\documentclass[fleqn,10pt]{wlscirep}
\usepackage{tabularx}
\usepackage{graphicx}
\usepackage{array}
\usepackage{bookmark}

\newcommand{\PreserveBackslash}[1]{\let\temp=\\#1\let\\=\temp}
\def\degree{${}^{\circ}$}
\title{Superconductivity at 33 - 37 K in $ALn_2$Fe$_4$As$_4$O$_2$ ($A$ = K and Cs; $Ln$ = Lanthanides)}

\author[1,+]{Si-Qi Wu}
\author[1,+]{Zhi-Cheng Wang}
\author[1,+]{Chao-Yang He}
\author[1]{Zhang-Tu Tang}
\author[1]{Yi Liu}
\author[1,2,3,*]{Guang-Han Cao}

\affil[1]{Department of Physics, Zhejiang University, Hangzhou
310027, China}
\affil[2]{State Key Lab of Silicon Materials, Zhejiang University, Hangzhou 310027, China}
\affil[3]{Collaborative Innovation Centre of Advanced Microstructures, Nanjing 210093, China}

\affil[*]{ghcao@zju.edu.cn}

\affil[+]{These authors contributed equally to this work}

\begin{abstract}
\textbf{We have synthesized 10 new iron oxyarsenides, K$Ln_2$Fe$_4$As$_4$O$_2$ ($Ln$ = Gd, Tb, Dy, and Ho) and Cs$Ln_2$Fe$_4$As$_4$O$_2$ ($Ln$ = Nd, Sm, Gd, Tb, Dy, and Ho), with the aid of lattice-match [between $A$Fe$_2$As$_2$ ($A$ = K and Cs) and $Ln$FeAsO] approach. The resultant compounds possess hole-doped conducting double FeAs layers, [$A$Fe$_4$As$_4$]$^{2-}$, that are separated by the insulating [$Ln_2$O$_2$]$^{2+}$ slabs. Measurements of electrical resistivity and dc magnetic susceptibility demonstrate bulk superconductivity at $T_\mathrm{c}$ = 33 - 37  K. We find that $T_\mathrm{c}$ correlates with the axis ratio $c/a$ for all 12442-type superconductors discovered. Also, $T_\mathrm{c}$ tends to increase with the lattice mismatch, implying a role of lattice instability for the enhancement of superconductivity.}

\end{abstract}
\begin{document}

\flushbottom
\maketitle
%
%
\thispagestyle{empty}

\noindent

\section*{Introduction}

\begin{figure}[ht]
\centering
\includegraphics[width=10cm]{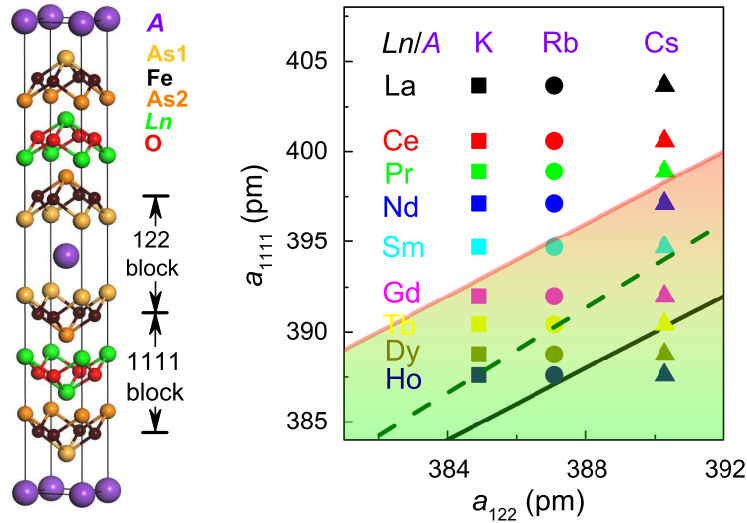}
\caption{Left: Crystal structure of 12442-type $ALn_2$Fe$_4$As$_4$O$_2$ ($A$ = K, Rb, and Cs; $Ln$ = Lanthanides), resulted from the intergrowth of 122-type $A$Fe$_2$As$_2$ and 1111-type $Ln$FeAsO. The right panel plots the lattice parameters of $A$Fe$_2$As$_2$ and $Ln$FeAsO\cite{LnO}. Each point ($a_{122}$, $a_{1111}$) denotes a 12442-type candidate. The lower and upper straight lines represent $a_{122}$ = $a_{1111}$ (an ideal match) and $a_{1111}$ = $a_{122}$ + 8 (the boundary of lattice match), respectively. The best lattice match is shifted to the middle dashed line due to the interlayer charge redistribution. The candidates in the shaded region are all successfully synthesized at ambient pressure.}
\label{lm}
\end{figure}

Since the discovery of superconductivity at 26 K in F-doped LaFeAsO in 2008\cite{hosono}, many compounds containing Fe\emph{X} (\emph{X}~=~As, Se) layers have been found to be superconducting at relatively high temperatures.\cite{jh,hosono-pc} These materials, known as Fe-based superconductors (FeSCs), have attracted tremendous research interests for their potential applications as well as for the rich underlying physics.\cite{hosono-pc,johnston,nsr}
By utilizing a structural-design strategy, we recently discovered the first double-FeAs-layer FeSC, KCa$_2$Fe$_4$As$_4$F$_2$, with a superconducting transition temperature $T_\mathrm{c}$ of 33 K.\cite{KCaF} KCa$_2$Fe$_4$As$_4$F$_2$ crystallizes in a so-called 12442-type structure resulting from an intergrowth between ThCr$_2$Si$_2$-type (122) and ZrCuSiAs-type (1111) blocks, as shown on the left of Fig.~\ref{lm}. Note that the 122-block KFe$_2$As$_2$ is heavily hole doped (0.5 holes per Fe atom) while the 1111-block CaFeAsF is non-doped. Consequently, the 12442-type material by itself is hole doped at a level of 0.25 holes per Fe atom, which makes it superconducting without extrinsic doping. Additionally, there are two distinct As sites in the FeAs layer, similar to the case in 1144-type materials \emph{AeA}Fe$_4$As$_4$ ($Ae$ = Ca, Sr, Eu; $A$ = K, Rb, Cs).\cite{1144,Eu1144,RbEu1144,CsEu1144} The most prominent feature of the 12442 structure lies in its separate double FeAs layers, in analogy with the double CuO$_2$ layers in cuprate superconductors.

We were able to expand the 12442-type superconducting family by simple elemental substitutions among alkali metals, which yield two additional fluo-arsenide FeSCs, $A$Ca$_2$Fe$_4$As$_4$F$_2$ with $A$ = Rb and Cs.\cite{ACaF} The $T_\mathrm{c}$ values are 30.5 K and 28.2 K, respectively. Then we also succeeded in synthesizing the first 12442-type oxyarsenide, RbGd$_2$Fe$_4$As$_4$O$_2$, whose $T_\mathrm{c}$ achieves 35 K.\cite{RbGd} With the substitution of rare-earth elements, four additional oxyarsenides in the Rb-containing series have been found.\cite{Rb_cm} These studies show that the lattice match between 122 and 1111 blocks is important, as suggested in our earlier proposal.\cite{jh} For this reason, the lattice-match paradigm can be employed for the exploration of the remaining 12442-type members. Fig.~\ref{lm} plots the combination of lattice parameters $a$ of \emph{A}Fe$_2$As$_2$ and \emph{Ln}FeAsO, ($a_{122}$, $a_{1111}$), each of which denotes a 12442-type candidate. The apparently ``ideal" lattice match is represented by the lower straight line with $a_{1111}$ = $a_{122}$. In the Rb-containing series (the middle column), the boundary for formation of the 12442 phase is at $Ln$ = Sm, corresponding to the lattice mismatch, parameterized as $2(a_{1111}-a_{122})/(a_{1111}+a_{122})$, of $\sim$2\%.\cite{Rb_cm} Therefore, we draw the upper-limit line, $a_{1111}$ = $a_{122}$ + 8 (pm), which could serve as a reference for exploration of the remaining 12442-type oxyarsenides.

In this paper, we report 10 new oxyarsenide FeSCs, K$Ln_2$Fe$_4$As$_4$O$_2$ ($Ln$ = Gd, Tb, Dy, and Ho) and Cs$Ln_2$Fe$_4$As$_4$O$_2$ ($Ln$ = Nd, Sm, Gd, Tb, Dy, and Ho), which coincidentally locate within the shaded area in Fig.~\ref{lm}. Other potential 12442 oxyarsenides outside the shaded area, such as KSm$_2$Fe$_4$As$_4$O$_2$, RbNd$_2$Fe$_4$As$_4$O$_2$ and CsPr$_2$Fe$_4$As$_4$O$_2$, cannot be synthesized at ambient pressure, which further confirms the crucial role of lattice match for the formation of the intergrowth structure. All these new commpounds synthesized show bulk superconductivity with $T_\mathrm{c}$ = 33 - 37 K. The maximum $T_\mathrm{c}$ of 37 K is observed for KGd$_2$Fe$_4$As$_4$O$_2$. The possible crystal-structure dependence of $T_\mathrm{c}$ is discussed in terms of the axis ratio and the lattice mismatch.

\section*{Results and Discussion}

\begin{figure}[ht]
\centering
\includegraphics[width=\linewidth]{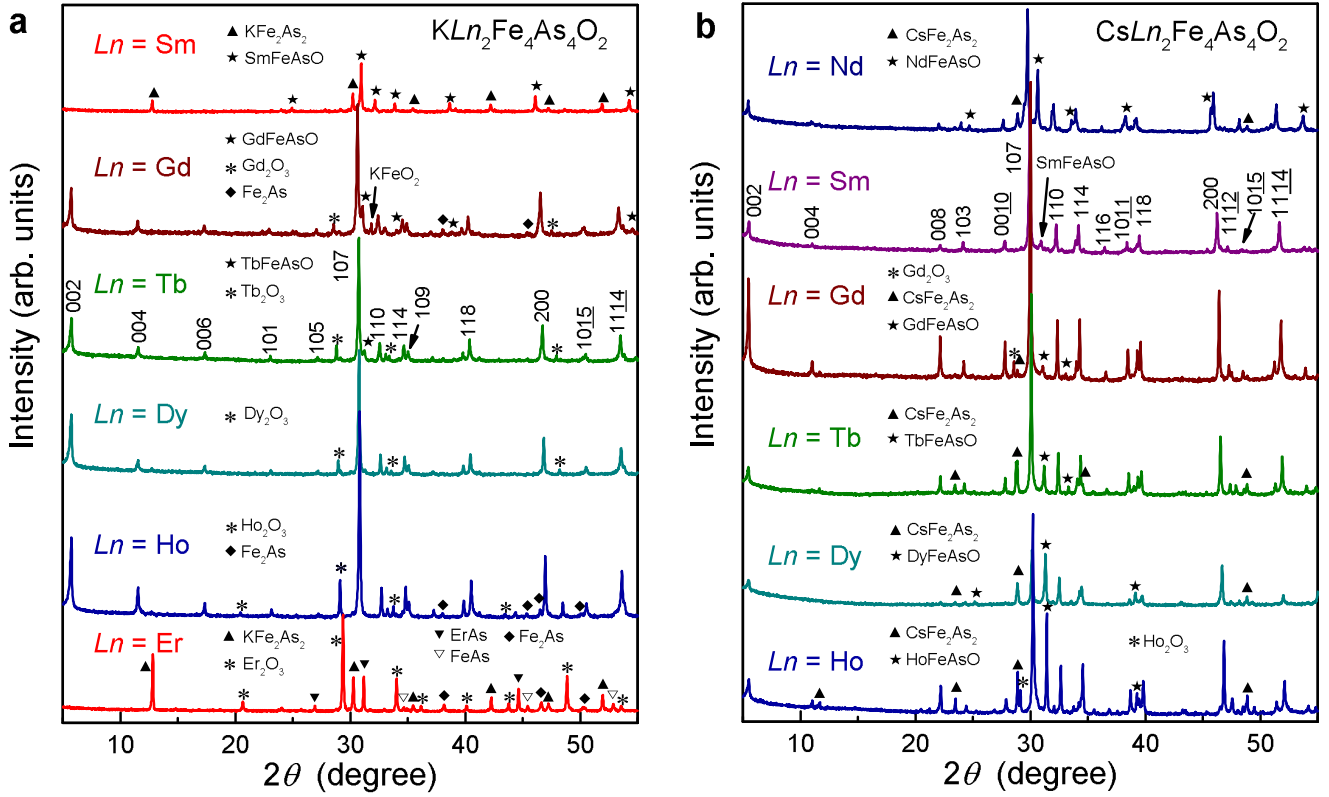}
\caption{Powder X-ray diffraction patterns for K$Ln_2$Fe$_4$As$_4$O$_2$ ($Ln$ = Sm, Gd, Tb, Dy, Ho, and Er) and Cs$Ln_2$Fe$_4$As$_4$O$_2$ ($Ln$ = Nd, Sm, Gd, Tb, Dy, and Ho) at room temperature. The indexed peaks are from the 12442 phase. Other peaks are from impurity phases which are labelled with different symbols as shown.}
\label{xrd}
\end{figure}

\textbf{X-ray Diffractions.} With the consideration of lattice match above, samples with the nominal composition of $ALn_2$Fe$_4$As$_4$O$_2$ ($A$ = K and Cs; $Ln$ = Nd, Sm, Gd, Tb, Dy, Ho, and Er) were prepared using high-temperature solid-state reactions (for details see the Method section). Fig.~\ref{xrd} shows the powder X-ray diffraction (XRD) results. In the series of $A$ = K (Fig.~\ref{xrd}a), the main XRD reflections for $Ln$ = Gd, Tb, Dy, and Ho can be indexed with a body-centered tetragonal lattice of $a \approx$ 3.88 {\AA} and $c \approx$ 30.6 {\AA}, consistent with the 12442-type structure. For $Ln$ = Sm, however, no reflections can be identified to the 12442 phase. The final products in the ``KSm$_2$Fe$_4$As$_4$O$_2$" sample are actually KFe$_2$As$_2$ and SmFeAsO. This is in contrast with the successful synthesis of RbSm$_2$Fe$_4$As$_4$O$_2$,\cite{Rb_cm} which clearly indicates the thermodynamic instability of KSm$_2$Fe$_4$As$_4$O$_2$ owing to its lattice mismatch (since it locates outside the shaded area in Fig.~\ref{lm}). For $Ln$ = Er, no 12442 phase appears either, although the lattice match seems desirable by extrapolation. Note that in this case the final product is a mixture of KFe$_2$As$_2$, Er$_2$O$_3$, ErAs, FeAs, and Fe$_2$As. The absence of ErFeAsO in the sample suggests that the failure in obtaining KEr$_2$Fe$_4$As$_4$O$_2$ is mainly due to the instability of ErFeAsO slabs.

The XRD patterns for the $A$ = Cs series are shown in Fig.~\ref{xrd}b. The main phase can be identified as the target 12442-type compounds with $Ln$ = Nd, Sm, Gd, Tb, Dy, and Ho. Similarly, syntheses of ``CsPr$_2$Fe$_4$As$_4$O$_2$" and ``CsEr$_2$Fe$_4$As$_4$O$_2$" failed because of the lattice mismatch and the instability of ErFeAsO, respectively. Note that the samples with $Ln$ = Nd, Tb, Dy, and Ho contain considerable amount of impurities, although optimization of the synthesis was performed. The impurities are mostly 122 and 1111 phases. This fact suggests that the stability of these compounds is marginal. Namely, the formation energy of the reaction, CsFe$_2$As$_2$ + 2 $Ln$FeAsO $\rightarrow$ Cs$Ln_2$Fe$_4$As$_4$O$_2$, is nearly zero especially for $Ln$ = Nd and Ho, which ultimately comes from the lattice mismatch.


Remarkably, both KHo$_2$Fe$_4$As$_4$O$_2$ and CsHo$_2$Fe$_4$As$_4$O$_2$ are realized, although HoFeAsO alone cannot be synthesized at ambient pressure. This means that the 12442-type structure may even stabilize the 1111 block. Interestingly, the HoFeAsO phase appears as the secondary phase in the CsHo$_2$Fe$_4$As$_4$O$_2$ sample. We conjecture that it is due to the decomposition of the 12442 phase in a certain high-temperature window.

\begin{figure}[ht]
\centering
\includegraphics[width=\linewidth]{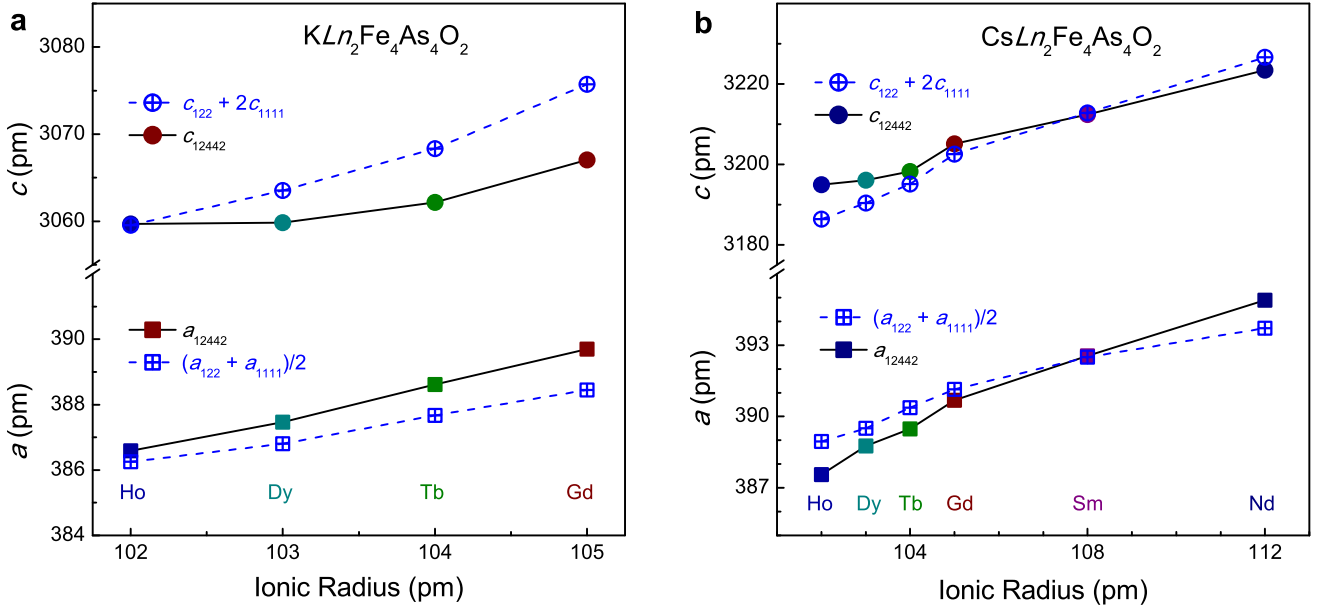}
\caption{Lattice parameters of K$Ln_2$Fe$_4$As$_4$O$_2$ and Cs$Ln_2$Fe$_4$As$_4$O$_2$ versus ionic radii of $Ln^{3+}$. The blue symbols with dashed lines denote the estimated values of the cell parameters from their constituent 122- and 1111-type unit cells.}
\label{cell}
\end{figure}

The lattice parameters can be calculated by using a least-squares fit, as tabulated in Tables~\ref{tab_K} and ~\ref{tab_Cs}. Fig.~\ref{cell}a-b plot the resultant lattice parameters of K$Ln_2$Fe$_4$As$_4$O$_2$ and Cs$Ln_2$Fe$_4$As$_4$O$_2$, respectively, as functions of the ionic radii of $Ln^{3+}$. The unit-cell parameters increase with the ionic radius of $Ln^{3+}$, as expected. To investigate the lattice-match effect, we also plot the estimated cell parameters, $(a_{122} + a_{1111})/2$ and $c_{122} + 2c_{1111}$, for comparison. Indeed, the estimated values basically agree with the experimental results. One would expect $a \approx a_{122} \approx a_{1111} \approx (a_{122} + a_{1111})/2$ for an ``ideal" lattice match. However, the best coincidence is seen for KHo$_2$Fe$_4$As$_4$O$_2$ and CsSm$_2$Fe$_4$As$_4$O$_2$ where the lattice match is apparently not perfect. This is due to the interlayer charge transfer which decreases $a_{1111}$, and simultaneously, increases $a_{122}$. Similar phenomenon is observed in the Rb$Ln_2$Fe$_4$As$_4$O$_2$ series in which the best lattice match is given by $Ln$ = Tb.\cite{Rb_cm} We thus draw the dashed line in the shaded area of Fig.~\ref{lm}, which marks the best lattice match after the charge-redistribution effect is taken into considerations. Above the dashed line, $a_{1111}$ is significantly larger than $a_{122}$, which means that the 122 (1111) block is under stretching. The resultant $a$ axis is larger than $(a_{122} + a_{1111})/2$. Below the dashed line, in contrast, the result is the opposite, which gives the crossings of data in Fig.~\ref{cell}b. Combining both results together, one concludes that the 122 block is more flexible to accommodate the lattice match. This conclusion is quite reasonable because the 1111 block contains relatively ``rigid" $Ln_2$O$_2$ layers.

\begin{table}[ht]
\renewcommand\arraystretch{2}
\centering
\begin{tabularx}{.88\textwidth}{p{.15\textwidth}<{\centering}p{.15\textwidth}<
{\centering}p{.15\textwidth}<{\centering}p{.15\textwidth}<{\centering}p{.15\textwidth}<{\centering}}
\hline\hline
$Ln$   & Gd & Tb & Dy & Ho  \\
\hline
$a$ ({\AA}) & 3.8970(0) & 3.8861(5) & 3.8746(3) & 3.8658(6)  \\
$c$ ({\AA}) & 30.670(3) & 30.621(8) & 30.598(4) & 30.597(1) \\
\hline\hline
\end{tabularx}
\caption{Lattice parameters of K\emph{Ln}$_2$Fe$_4$As$_4$O$_2$ (\emph{Ln}~=~Gd, Tb, Dy, and Ho) at room temperature.}
\label{tab_K}
\end{table}

\begin{table}[ht]
\renewcommand\arraystretch{2}
\centering
\begin{tabularx}{.88\textwidth}{p{.1\textwidth}<{\centering}p{.1\textwidth}<
{\centering}p{.1\textwidth}<{\centering}p{.1\textwidth}<{\centering}p{.1\textwidth}<
{\centering}p{.1\textwidth}<{\centering}p{.1\textwidth}<{\centering}}
\hline\hline
$Ln$  & Nd & Sm & Gd & Tb & Dy & Ho  \\
\hline
$a$ ({\AA}) & 3.9488(4) & 3.9255(6) & 3.9067(7) & 3.8948(1) & 3.8876(6) & 3.8756(2)  \\
$c$ ({\AA}) & 32.234(1) & 32.123(8) & 32.051(3) & 31.982(3) & 31.960(8) & 31.949(1) \\
\hline\hline
\end{tabularx}
\caption{Lattice parameters of Cs\emph{Ln}$_2$Fe$_4$As$_4$O$_2$ (\emph{Ln}~=~Nd, Sm, Gd, Tb, Dy, and Ho) at room temperature.}
\label{tab_Cs}
\end{table}

\

\begin{figure}[ht]
\centering
\includegraphics[width=\linewidth]{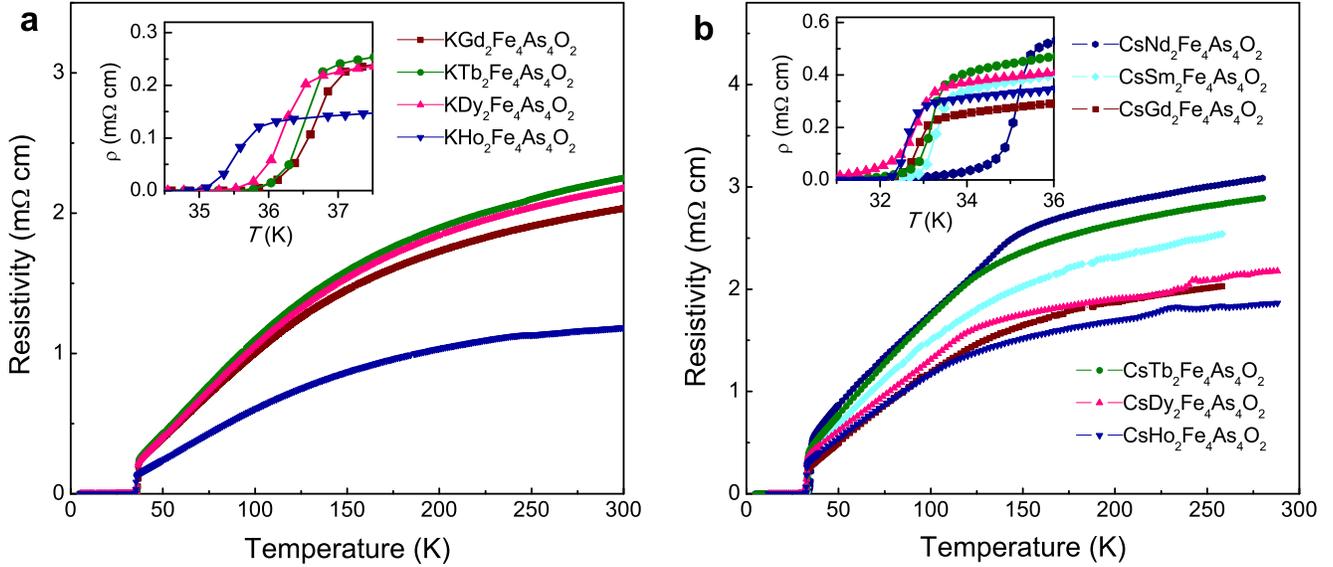}
\caption{Temperature dependence of resistivity for the K$Ln_2$Fe$_4$As$_4$O$_2$ (left panel) and Cs$Ln_2$Fe$_4$As$_4$O$_2$ (right panel) polycrystalline samples. The insets zoom in the superconducting transitions.}
\label{RT}
\end{figure}

\textbf{Superconductivity.} Fig.~\ref{RT} shows the temperature dependence of resistivity, $\rho(T)$, for K$Ln_2$Fe$_4$As$_4$O$_2$ ($Ln$ = Gd, Tb, Dy, and Ho) and Cs$Ln_2$Fe$_4$As$_4$O$_2$ ($Ln$ = Nd, Sm, Gd, Tb, Dy, and Ho). The $\rho(T)$ data indicate a metallic conduction with a negative curvature in the high-temperature region. The obvious negative curvature is different from the linear behaviour expected for a dominant electron-phonon scattering. This phenomenon is very often seen in hole-doped FeSCs\cite{Ba122,whh,KCaF}, which could be due to an incoherent-to-coherent crossover\cite{Kondo}. Note that the kink-like feature in Cs$Ln_2$Fe$_4$As$_4$O$_2$ ($Ln$ = Nd, Dy and Ho) comes from the SDW anomalies of the secondary phases $Ln$FeAsO. Another prominent feature is the linear $\rho(T)$ below $\sim$80 K in the normal state, which contrasts with the conventional $T^n$ ($n\geq$2) behaviour due to electron-phonon and/or electron-electron scatterings, suggesting a non-Fermi-liquid behaviour.

Superconducting transitions appear at $T_{\mathrm{c}}^{\mathrm{onset}}$ = 36.0 - 37.1 K for K\emph{Ln}$_2$Fe$_4$As$_4$O$_2$, and 32.9 - 35.4 K for Cs\emph{Ln}$_2$Fe$_4$As$_4$O$_2$. The transition widths (the difference in temperature at which the resistivity drops to 90\% and 10\% of the extrapolated normal-state value) are typically about 0.5 K, albeit of a transition ``tail" for Cs\emph{Ln}$_2$Fe$_4$As$_4$O$_2$ ($Ln$ = Nd and Dy) which contain relatively large amount of non-superconducting secondary phases. $T_\mathrm{c}$ decreases monotonically with the increase of the atomic number of the lanthanides, akin to the case of Rb-containing series. Note that, in the latter system, the $T_\mathrm{c}$ value correlates with the normal-state resistivity.\cite{Rb_cm} Similar trends can also be seen in the present two systems, even though the samples' quality makes some interferences. Since the normal-state resistivity is not dominated by the electron-phonon scattering, as stated above, the $T_\mathrm{c}$ correlation suggests non-electron-phonon mechanisms for the superconductivity.

\begin{figure}[ht]
\centering
\includegraphics[width=\linewidth]{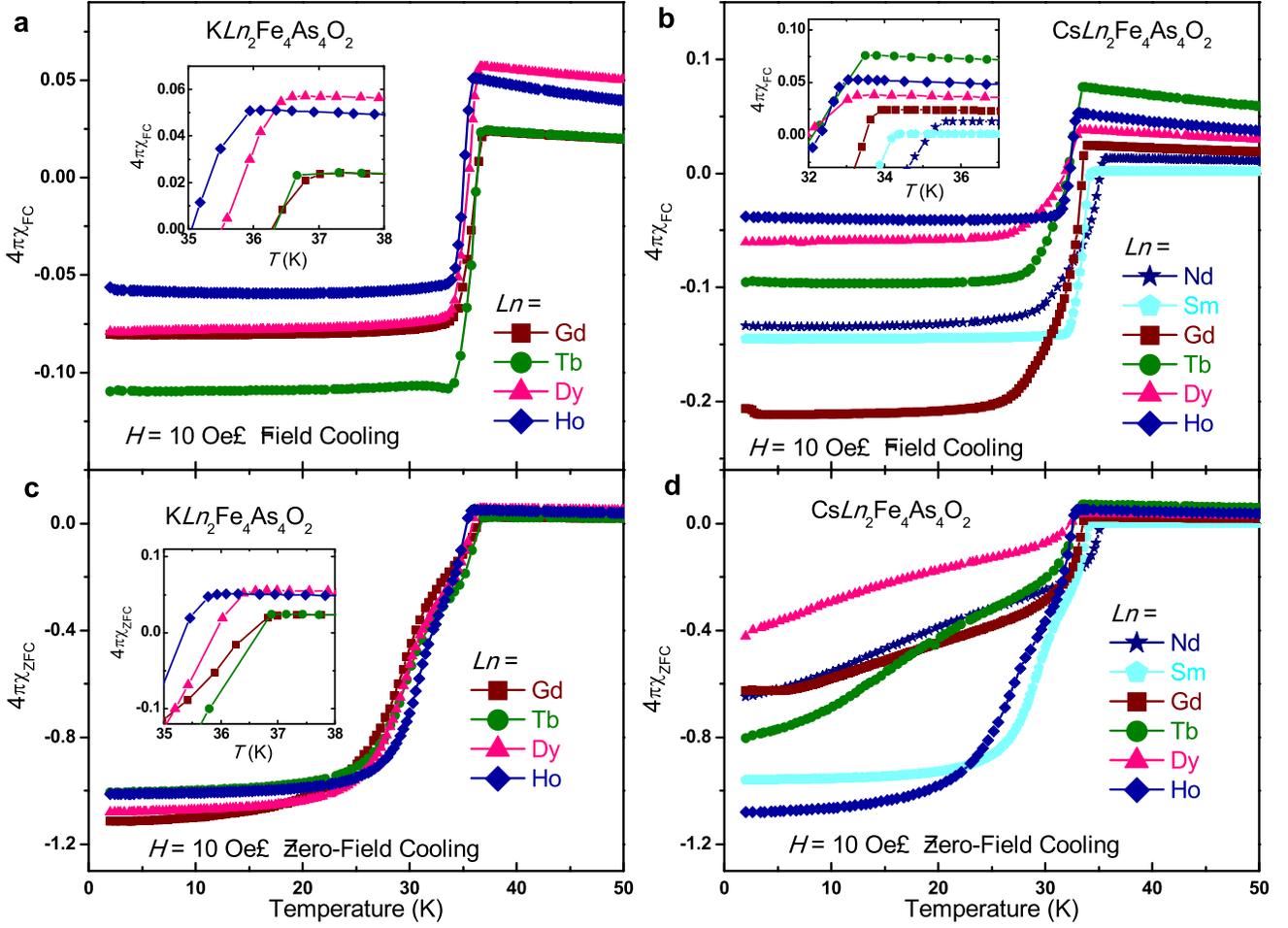}
\caption{Temperature dependence of the magnetic susceptibility (in $4\pi\chi$) for K\emph{Ln}$_2$Fe$_4$As$_4$O$_2$ (left) and Cs\emph{Ln}$_2$Fe$_4$As$_4$O$_2$ (right). Both field-cooling (upper panels) and zero-field-cooling (bottom panels) data are presented. The insets zoom in the superconducting transitions.}
\label{Mag}
\end{figure}

Superconductivity in K$Ln_2$Fe$_4$As$_4$O$_2$ and Cs$Ln_2$Fe$_4$As$_4$O$_2$ is verified by the dc magnetic measurements. Fig.~\ref{Mag} shows the temperature dependence of magnetic susceptibility ($\chi$). For clearness, the data in the field-cooling (FC) and zero-field-cooling (ZFC) protocols are plotted separately. First of all, $\chi_{\mathrm{FC}}$ decreases abruptly at 33 - 37 K owing to the superconducting Meissner effect. Note that the normal-state susceptibility is mainly contributed from the Curie-Weiss-type paramagnetism of the lanthanide moments. The Meisnner volume fraction, estimated from the magnitude of the $\chi_{\mathrm{FC}}$ drop, is typically about 10\% for the samples with relatively less impurities. Nevertheless, the superconducting shielding fractions, shown in Fig.~\ref{Mag}c and d, are several times larger. The reduced superconducting signal in FC mode is very often seen for type-II superconductors (including FeSCs), because of the magnetic-flux pinning when cooling under magnetic fields. For polycrystalline samples, the magnetic shielding fractions better indicate the superconducting volume fraction. The step-like anomaly below $T_\mathrm{c}$ for the ZFC data, which also frequently appears for polycrystalline samples of extremely type-II
superconductors, comes from the intergrain weak links. As none of the impurities shows superconductivity above 5 K (only KFe$_2$As$_2$ and CsFe$_2$As$_2$ are superconducting at $T_\mathrm{c}$ = 3.8~K and 2.6~K, respectively\cite{A122}), we conclude that the title compounds are responsible for the superconductivity.

\
\textbf{Discussion.} Above we demonstrate that all the 10 new 12442-type compounds show bulk superconductivity at $T_\mathrm{c}$ = 33 - 37 K. In fact, the variation in $T_\mathrm{c}$ spans nearly 10 K ( from 28.2 to 37 K), if the previously discovered 12442-type FeSCs are included. Note that, for a specific 12442-type superconductor, the $T_\mathrm{c}$ value hardly changes regardless of the sample's purity. This means that $T_\mathrm{c}$ is intrinsically determined by the material itself, primarily because the material is a line compound (with the stoichiometric composition) which gives rise to a constant hole doping (0.25 holes/Fe-atom). We also note that the magnetism of the lanthanide hardly influences the $T_\mathrm{c}$ value.\cite{Rb_cm} Therefore, it is meaningful to examine the crystal-structure dependence of $T_\mathrm{c}$. Our previous investigations\cite{ACaF,Rb_cm} show that the $T_\mathrm{c}$ trend does not follow the well-known empirical rules: the As$-$Fe$-$As bond angle\cite{zhaoj,lee} and the As height from the Fe plane\cite{mizuguchi} are the two crucial parameters for determining $T_\mathrm{c}$. The possible reason is that the 12442-type FeSCs exceptionally contain double FeAs layers. We found that $T_\mathrm{c}$ decreases (increases) with the intra-bilayer (inter-bilayer) spacing.\cite{ACaF,Rb_cm} In the present K- and Cs-containing series, similar behaviour is expected (here we cannot give the structural correlation plot because some samples' purity is not good enough for a reliable structural refinement).

\begin{figure}[ht]
\centering
\includegraphics[width=\linewidth]{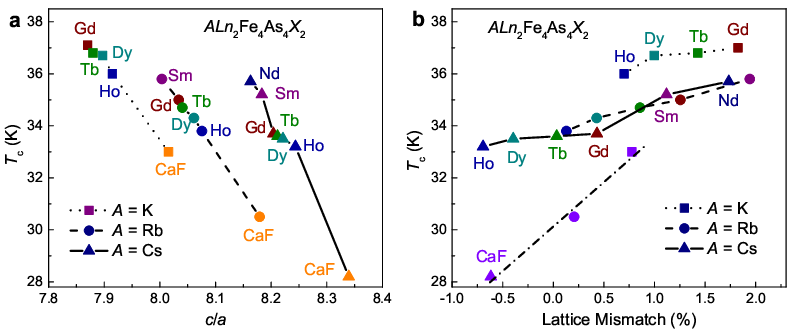}
\caption{Influence of $c/a$ (left panel) and the lattice mismatch, $2(a_{1111}-a_{122})/(a_{1111}+a_{122})$, (right panel) on $T_\mathrm{c}$ for all the 12442-type superconductors. $A$, $Ln$, and $X$ denote alkali metals, lanthanides, and oxygen or fluorine, respectively.}
\label{Tc}
\end{figure}

The intra-bilayer and inter-bilayer spacings are defined as the thickness of 122 and 1111 block layers (see Fig.~\ref{lm}).\cite{ACaF,Rb_cm} For a certain series, e.g. $A$ = K, while the inter-bilayer spacing definitely increases with the lanthanide-ion radius, the intra-bilayer spacing is actually influenced by the lattice mismatch. As we mentioned above, the 122 block layer is more flexible with respect to the lattice mismatch. If the 122 bock is under stretching, intra-bilayer spacing is decreased, and vice versa. This means that the axis ratio, $c/a$, which is easily accessible, may serve as an effective parameter instead of the intra-bilayer spacing. Fig.~\ref{Tc}a plots $T_\mathrm{c}$ vs. $c/a$ for all 12442-type superconductors. One sees that $T_\mathrm{c}$ decreases almost linearly with $c/a$ for each series with $A$ = K, Rb, and Cs. The result suggests that the intra-bilayer coupling could enhance the superconductivity.

Note that the change of intra-bilayer spacing is resulted from the lattice mismatch, as discussed above, we also plot $T_\mathrm{c}$ vs. $2(a_{1111}-a_{122})/(a_{1111}+a_{122})$. As shown in Fig.~\ref{Tc}b, $T_\mathrm{c}$ basically increases with the lattice mismatch. The result is actually consistent with the axis-ratio dependence, since a positive lattice mismatch gives rise to a small $c/a$. Nevertheless, the lattice mismatch also marks the lattice instability. This implies that lattice instability might play a role for the $T_\mathrm{c}$ enhancement.

So far the highest $T_\mathrm{c}$ value is seen in KGd$_2$Fe$_4$As$_4$O$_2$ (37 K) among all the eighteen 12442-type FeSCs discovered. It is of great interest whether the $T_\mathrm{c}$ value may exceed the record of hole-doped FeSCs (38 K in Ba$_{0.6}$K$_{0.4}$Fe$_2$As$_2$\cite{Ba122}), since more 12442-type compounds are likely to be discovered. As we previously pointed out, 12442-type compounds with the incorporation of actinides, such as Rb$At_2$Fe$_4$As$_4$O$_2$ ($At$ = Np and Pu), seem to be synthesizable at ambient pressure.\cite{Rb_cm} Furthermore, additional metastable 12442-type iron pnictides (say, those outside of the shaded area in Fig.~\ref{lm}) could also be realized by high-pressure synthesis.

\section*{Concluding Remarks}
In summary, we report the syntheses, structural characterisations and physical-property measurements of 10 quinary iron arsenides, $ALn_2$Fe$_4$As$_4$O$_2$ ($A$ = K and Cs; $Ln$ = Lanthanides). The new materials consist of double FeAs layers that are intrinsically hole-doped (0.25 holes/Fe-atom), which give rise to the emergence of bulk superconductivity at 33 - 37 K. The lanthanide-ion magnetism hardly influence the superconductivity. Owing to the unique double-FeAs-layer structure, an additional structural parameter, the bilayer thickness, is found to correlate with the $T_\mathrm{c}$ value. On the other hand, the lattice mismatch seems to enhance the superconductivity also.

We confirm that the lattice match between the constituent blocks is crucial to stabilize the 12442-type intergowth structure. Now there have been 18 members in the 12442 family, nevertheless, additional 12442-type FeSCs are hopefully to be discovered, in particular, via a high-pressure synthesis.

\section*{Methods}

Samples of \emph{A}\emph{Ln}$_2$Fe$_4$As$_4$O$_2$ (\emph{Ln}~=~Gd, Tb, Dy, Ho for \emph{A}~=~K; \emph{Ln}~=~Nd, Sm, Gd, Tb, Dy, Ho for \emph{A}~=~Cs) were synthesized via solid state reaction processes similar to the previously reported KCa$_2$Fe$_4$As$_4$F$_2$\cite{RbGd}. First, source materials of Fe powder (99.998\%) and \emph{Ln} powder (99.9\%) were respectively mixed with As granule (99.999\%), sealed in evacuated quartz tubes, and heated to 750~\degree C to prepare FeAs, Fe$_2$As, and \emph{Ln}As. Intermediate product of ``\emph{A}$_{1.03}$Fe$_2$As$_2$'' was then produced by reacting FeAs and \emph{A} (99.5\%) at 600 - 650~\degree C for 10 h. After that, stoichiometric mixtures of \emph{A}$_{1.03}$Fe$_2$As$_2$, FeAs, Fe$_2$As, \emph{Ln}As, and lanthanide oxides (pre-heated to 900~\degree C for 24 h to remove adsorbed water) were ground, pelletized, and finally sintered for 36~h at 940 - 970~\degree C in alumina tubes which were sealed in Ta tubes jacketed with evacuated quartz ampoules. The final products were found to be stable in air.

We employed a PANalytical X-ray diffractometer with Cu K$\alpha$1 radiation to conduct the powder X-ray diffraction experiments at room temperature. The electrical resistivity measurements were carried out via a standard four-probe method on a Physical Property Measurement System (PPMS-9, Quantum Design), with a excitation current of 2~mA. Magnetic properties were measured on a Magnetic Property Measurement System (MPMS-XL5) under a magnetic field of 10~Oe. The samples were cut and polished into rods, and the applied field is along the rod direction, which minimizes the demagnetization effect.

\bibliography{12442_SR}









\textbf{Acknowledgements}

This work was supported by the National Science Foundation of China (Nos. 11474252 and 11190023) and the National Key Research and Development Program of China (No. 2016YFA0300202).
\\

\textbf{Author Contributions}

G.H.C. coordinated the work. Z.C.W., S.Q.W. and C.Y.H. synthesized, characterized and measured the samples with the help from Z.T.T. and Y.L. The paper was written by G.H.C. and S.Q.W. All the authors reviewed the manuscript.
\\

\textbf{Additional Information}

Competing financial interests: The authors declare no competing financial interests.

\end{document}